\documentclass[aps,pre,showpacs,showkeys,preprint,floatfix,superscriptaddress]{revtex4-1}

\usepackage{amssymb}
\usepackage{amsmath}
\usepackage{amsbsy}
\usepackage{bm}
\usepackage{color}
\usepackage{graphicx}

\newcommand{\bq}{\begin{eqnarray}}
\newcommand{\eq}{\end{eqnarray}}
\newcommand{\bqn}{\begin{eqnarray*}}
\newcommand{\eqn}{\end{eqnarray*}}
\newcommand{\RR}{{\bf R}}
\newcommand{\rr}{{\bf r}}

\newcommand{\LL}{{\bf L}}

\newcommand{\pint}{\int^{\!\raisebox{-1.5pt}{\mbox{\textendash}}}_{\!\!\!\!\raisebox{-1pt}{\mbox{\textendash}}}\!\!\!\!\!\!\!\int}
\newcommand{\ZZ}{\sf{Z\!\!Z}}
\newcommand{\real}{I\!\!R}

\begin{document}
\title{How should we choose the boundary conditions in a
  simulation which could detect anyons in one and two dimensions?} 

\author{Riccardo Fantoni}
\email{rfantoni@ts.infn.it}
\affiliation{Universit\`a di Trieste, Dipartimento di Fisica, strada
  Costiera 11, 34151 Grignano (Trieste), Italy}

\date{\today}

\begin{abstract}
We discuss the problem of anyonic statistics in one and two
spatial dimensions from the point of view of statistical physics. In
particular we want to understand how the choice of the Born-von
Karman or the twisted periodic boundary conditions necessary in a
Monte Carlo simulation to mimic the thermodynamic limit of the many
body system influences the statistical nature of the particles. The
particles can either be just bosons, when the configuration space is
simply connected as for example for particles on a line. They can be
bosons and fermions, when the configuration space is doubly connected
as for example for particles in the tridimensional space or in a
Riemannian surface of genus greater or equal to one (on the torus, etc
\ldots). They can be scalar anyons with arbitrary statistics, when the
configuration space is infinitely connected as for particles on the
plane or in the circle. They can be scalar anyons with fractional
statistics, when the configuration space is the one of particles on a
sphere. One can further have multi components anyons with fractional 
statistics when the configuration space is doubly connected as for
particles on a Riemannian surface of 
genus greater or equal to one. We determine an expression for the
canonical partition function of hard core particles (including anyons)
on various geometries. We then show how the choice of boundary
condition (periodic or open) in one and two dimensions determine which
particles can exist on the considered surface. In the conclusion, we
mention the Laughlin wavefunction and give a few comments about
experiments. 
\end{abstract}



\maketitle
\section{Introduction}
\label{sec:introduction}

For the statistical mechanics of a systems of many anyons very partial
results can be obtained, because the exact solution of a gas of anyons
is not known. In fact, in contrast to the bosonic or fermionic case
where the statistics is implemented by hand on the many body Hilbert
space by constructing completely symmetric or antisymmetric products
of single particle wave functions, for anyons the complicated boundary
conditions for the interchange of any two particles require the
knowledge of the complete many-body configurations. Only the two-body
problem is exactly soluble for anyons, and hence only the two-body
partition function can be computed exactly. Since the thermodynamic
limit cannot be performed, one has to resort to approximate or
alternative methods to study the statistical mechanics of anyons
\cite{Ouvry2007,Stern2008}. For 
example if the thermodynamic functions are analytic in the particle
density, it is well-known that the low density, or equivalently the
high temperature limit, of a (free) gas can be investigated using the
virial expansion. 

Anyons have had important physical applications and it would be wrong
to convey the idea that they are just mathematical fantasies. For
example physical objects which can be described as anyons are the
quasi-particle and quasi-hole excitations of planar systems of
electrons 
exhibiting the  fractional quantum Hall effect (QHE) (for a review see
for instance \cite{Prange}). Most of the great interest that anyonic
theories have attracted in the past few years derives precisely from
their relevance to a better  understanding of the fractional QHE
\cite{Halperin1984}, in conjunction with several claims that anyons can
provide also a non-standard explanation of the mechanism of high
temperature superconductivity \cite{Chen1989}. Even if recent
experiments have cast some shadow on the relevance of fractional
statistics to the observed high temperature superconductivity
\cite{Lyons1990,Kiefl1990,Spielman1990}.   

In this work we focus on the important problem of how the boundary
conditions on the simulation box influences the statistics of the
anyonic (see chapter 2 of Ref. \cite{Lerda}) particles. We will
consider various cases: the infinite line, the circle, the infinite
plane, the torus, and the sphere. In each case we will determine the
nature of the statistics of the many anyons system. This is important
because in a simulation of a real material one usually chooses
periodic boundary conditions in order to approach the thermodynamic
limit.  

Another interesting problem is the determination of a spinor for an
anyon with a given rational or even irrational (either algebraic or
even transcendental) statistics. If the spin-statistics theorem
\cite{Pauli1940} which states that, as a consequence of Lorentz
invariance and of locality, half integer spin particles must obey to
Fermi statistics and integer spin particles must obey to Bose
statistics, there is nothing similar for anyonic statistics
\cite{Oeckl2001}. Citing Wilczek \cite{Wilczek90} we can say that
``The basic difficulty, which makes this problem much more difficult
for generic anyons than for bosons or fermions, is that for generic
anyons the many-body Hilbert space is in no sense the tensor product
of the one-particle Hilbert space. This circumstance can be understood
in various ways. Its root is that in the general case the weighting
supplied by anyon statistics depends not only on the initial and final
states, but also on a (topological) property of the trajectory
connecting them. This means that in the general case it is impossible
to summarize the effect of quantum statistics by projection on the
appropriate weighted states, as we do for bosons and fermions --
where, of course, we project respectively on symmetric and
antisymmetric states''. We will consider this problem in a future
work.

\section{The statistical physics anyon problem in two dimensions}
\label{sec:problem}

The statistical mechanical properties of a quantum system of $N$ hard
core particles in a volume $V$ in $d$ spatial dimensions occupying
positions $q\in (\real^d)^N$ and described by an Hamiltonian
$\hat{{\cal H}}$ in thermal equilibrium at the inverse temperature
$\beta=1/k_BT$, with $k_B$ the Boltzmann constant and $T$ the absolute
temperature, are obtainable from the thermal density matrix operator
\cite{Wu1984},   
\bq
\hat{\rho}=\exp(-\beta\hat{{\cal H}}).
\eq

In the configurations space representation the thermal density matrix
can be written using the following path integral notation,
\bq \label{dm}
\rho(q^\prime,q;\beta)=\sum_{\alpha\in\pi_1(M_N^d)}\chi(\alpha)
\pint_{q_\alpha(0)=q}^{q_\alpha(\hbar\beta)=q^\prime} e^{
-\frac{1}{\hbar}\int_0^{\hbar\beta}d\tau\,{\cal H}(q_\alpha(\tau),
\dot{q}_\alpha(\tau))}\, {\cal D}q_\alpha,
\eq

where ${\cal H}(q,\dot{q})$ is the classical Hamiltonian of the $N$ 
hard core, identical particles. The meaning of $\pi_1(M_N^d)$ and of
the phases $\chi$ will be shown in the next two sections. 

The canonical partition function can then be found from the trace of
the density matrix,
\bq
Z(N,V,T)=\int \rho(q,q;\beta)\,dq. 
\eq
\subsection{$M_N^d$ and its fundamental group}
Consider a system of $N$ identical hard core particles moving in the
euclidean $d$-dimensional space, $\real^d$. A configuration of such a
system is clearly specified by the $N$ coordinates of the particles,
i.e. by an element of $(\real^d)^N$. However because of the hard core
assumption, any two particles cannot occupy the same position. So from
$(\real^d)^N$ we have to remove the diagonal,
\bq
\Delta=\{(\rr_1,\ldots,\rr_N)\in (\real^d)^N :\rr_i=\rr_j ~\mbox{for some}
~i\ne j\}.
\eq
Furthermore our particles are identical and indistinguishable, so we 
should identify configurations which differ only in the ordering of the
particles. In other words we should divide by the permutation group
$S_N$. Therefore we conclude that the configuration space for our system is
\bq
M_N^d=\frac{(\real^d)^N-\Delta}{S_N}.
\eq
To find the fundamental group of such space is a standard problem in
algebraic topology, which was solved in the early 60's
\cite{Fadell1962,Fox1962,Fadell1962b}.  
It turns out that the fundamental group of $M_N^d$ is given by
\bq \label{braid}
\pi_1(M_N^d)=\left\{ 
\begin{array}{c}
S_N~~\mbox{if} ~~d\ge 3\\
B_N~~\mbox{if} ~~d=2
\end{array}\right.
\eq
where $B_N$ is Artin' s braid group of $N$ objects which has the
permutation group $S_N$ as a homomorphic image
\cite{Artin1926,Artin1947}.

\begin{figure}[htbp]
\begin{center}
\includegraphics[width=8cm]{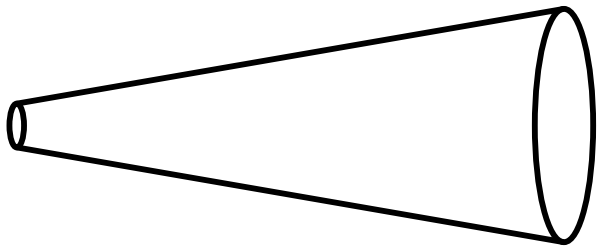}
\end{center}  
\caption{Schematic representation of $r_2^2$ with the topology of a
  cone without the tip. It is an infinitely connected space.} 
\label{fig:2d}
\end{figure}

Even from this formal point of view we see that there is a crucial 
difference between two and three or more dimensions. To have a more
explicit understanding of (\ref{braid}), let us consider a two
particle example in the light of what we have just observed. Let us
start with the case of two dimensions. Instead of assigning the
position vectors $\rr_1$ and $\rr_2$ for the two particles, is more
convenient to introduce the center of mass coordinate,
\bq
\RR=\frac{1}{2}(\rr_1+\rr_2)~~\in \real^2,
\eq
and the relative coordinate,
\bq
\rr=\rr_1-\rr_2~~\in \real^2-\{0\}.
\eq
We have removed the origin because of the hard core requirement. Since  
$\RR$ is invariant under the permutations of $S_2$, we can write,
\bq \label{22}
M_2^2=r_2^2\times\real^2,
\eq 
where $r_2^2$ is some space describing the two degrees of freedom of the 
relative motion. We now argue that $r_2^2$ has the topology of a cone.
Since two configurations which differ only in the ordering of the 
particle indexes are indistinguishable, $\rr$ and $-\rr$ must be 
identified. The space $r_2^2$ is then the upper half plane without 
the origin and with the positive x-axis identified with the negative one,
i.e. is a cone without the tip (see fig. \ref{fig:2d}).

According to the decomposition (\ref{22}), any loop in $M_2^2$ can be 
classified by the number of times it winds around the cone $r_2^2$.
Two loops $q$ and $q^\prime$ with different winding numbers are 
homotopically inequivalent: it is not possible to deform one into 
the other since the vertex of the cone has been removed. Thus the 
space $r_2^2$ and $r_2^2\times\real^2$, are infinitely connected,
and,
\bq
\pi_1(M_2^2)=\pi_1(r_2^2\times\real^2)=\ZZ=B_2.
\eq

It is important to realize that if the vertex of the cone were included
(i.e. allowing particles to occupy the same position in space) the 
configuration space would be simply connected. Any loop, even when winding 
around the cone, would be contracted to a point by deforming and 
unwinding it through the tip. Thus, if we do not impose the hard core 
constraint on the particles, we can describe only bosonic statistics.

\begin{figure}[htbp]
\begin{center}
\includegraphics[width=8cm]{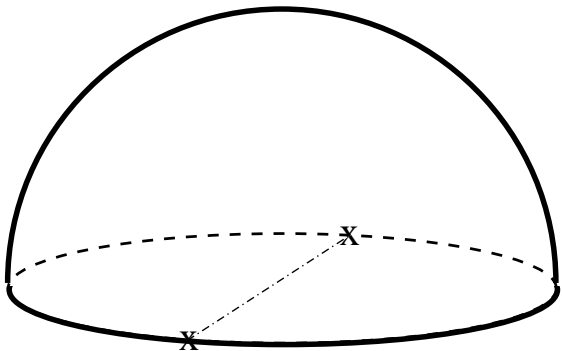}
\end{center}  
\caption{Schematic representation of ${\cal  P}_2$ as the northern
  hemisphere with opposite points on the equator being identified. It
  is a doubly connected space.} 
\label{fig:3d}
\end{figure}

Let us now turn to the case of two particles in three dimensions. After
introducing the center of mass coordinate $\RR\in\real^3$, we can decompose 
the configurations space as,
\bq \label{23}
M_2^3=r_2^3\times\real^3,
\eq
where the space $r_2^3$ describes the three degrees of freedom of the 
relative motion. These are the length and the two angles of the relative
coordinate $\rr$. As before $\rr$ and $-\rr$ are identified. It is easy to 
realize that $r_2^3$ is just the product of the semi-infinite line 
describing $|\rr|$ and the projective space ${\cal  P}_2$ describing the
orientation of $\pm\rr/|\rr|$. In turn ${\cal  P}_2$ can be described 
as the northern hemisphere with opposite points on the equator being 
identified (see fig. \ref{fig:3d}).

\begin{figure}[htbp]
\begin{center}
\includegraphics[width=8cm]{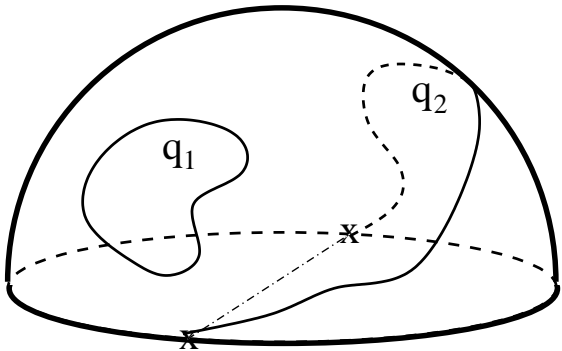}
\end{center}  
\caption{Schematic representation of a a typical contractible loop
  $q_1$ and a typical non-contractible loop $q_2$ on $M_2^3$ which is
  a doubly connected space. The two points marked with an $X$ are the
  same point. The path $q_1$ cannot be deformed continuously into the
  path $q_2$.}  
\label{fig:3ddc}
\end{figure}

The space ${\cal  P}_2$ is doubly connected and admits two classes of loops:
those which can be shrunk to a point by a continuous transformation and 
those which cannot. In fig. \ref{fig:3ddc} we exhibit a typical
contractible loop $q_1$ and a typical non-contractible loop $q_2$.

Therefore from the decomposition (\ref{23}) and the topology of $r_2^3$,
we deduce that,
\bq
\pi_1(M_2^3)=\pi_1(r_2^3\times\real^3)=\ZZ_2=S_2.
\eq
Thus only bosons and fermions can exist, the former corresponding to
contractible loops and the latter to non-contractible loops.

We have seen that at the heart of the anyonic statistics there is the 
braid group $B_N$ in place of the permutation group $S_N$ which is 
responsible for ordinary statistics. There are only two one dimensional
unitary representations 
of $S_N$, namely the identical one, $\chi(\mbox{even and odd 
permutations})=+1$  (bosonic statistics) and the alternating one, 
$\chi(\mbox{even permutations})=+1$, $\chi(\mbox{odd permutations})=-1$
(fermionic statistics). Whereas the braid group admits a whole 
variety of one dimensional 
\footnote{When dealing with non-scalar quantum mechanics, i.e. when the 
wave functions are multiplets instead of one component objects as assumed
in the discussion, appropriate higher dimensional representations of 
$\pi_1(M_N^d)$ would be necessary.}
unitary representations whose labeling parameter will be identified
with the parameter $\nu$ also called the {\sl statistics}.

\subsection{Statistical mechanics problem}
\label{sec:smp}
One is usually interested in calculating the partition function of the 
system which is given by the trace of the density matrix.
So we choose $q=q^\prime$, or loops in $M_N^d$. Two loops are
considered equivalent (or homotopic) if one can be obtained 
from the other by a continuous deformation. All homotopic loops are 
grouped into one class and the set of all such classes is called the 
{\sl fundamental group} and is denoted by $\pi_1$
\footnote{In the set $\pi_1$ one can define a product $\cdot$ in a very 
simple and natural way: if $\alpha_1$ and $\alpha_2$ are two classes
with representatives path $q_1$ and $q_2$, then $\alpha_1\cdot
\alpha_2$ is the class whose representative is the path $q_1q_2$
(that is the pat $q_1$ followed by the path $q_2$). It can be shown 
that this product furnishes $\pi_1$ with a group structure.}.
Thus an element of $\pi_1(M_N^d)$ is simply the set of all loops in 
$M_N^d$ which can be continuously deformed into each other.
On the other hand, loops belonging to two different elements of 
$\pi_1(M_N^d)$ cannot be connected by a continuous transformation.
Naturally $\rho(q^\prime,q;\beta)$ has to be a real positive probability
function. 

In order for (\ref{dm}) to make sense as a probability amplitude,
the complex weights $\chi(\alpha)$ cannot be arbitrary. In fact, since 
we want to maintain the usual rule for combining probabilities,
\bq
\rho(q^\prime,q;\beta)=\int_{M_N^d}dq_o\,\rho(q^\prime,q_o;t_o/\hbar)
\rho(q_o,q;\beta-t_o/\hbar)~~~,
\eq
the weights $\chi(\alpha)$ must satisfy,
\bq \label{rep}
\chi(\alpha_1)\chi^\star(\alpha_2)=\chi(\alpha_1\cdot\alpha_2)~~,
\eq
for any $\alpha_1$ and $\alpha_2$. Equation (\ref{rep}) can also be read 
as the statement that $\chi(\alpha)$ must be a one dimensional 
unitary ($|\chi|^2=1$) representation of the fundamental group $\pi_1(M_N^d)$
\cite{Laidlaw1971}. To see which representations are possible, we have
to specify better what is $M_N^d$ and its fundamental group.

This means that we have to look for one dimensional unitary
representations $\chi(\alpha)$ of the fundamental group, i.e. 
\bq
\chi(\alpha)=e^{-i\nu n_\alpha \pi},~~~n_\alpha~~\mbox{integer},
\eq
or in the notation used by F. Wilczek \cite{Wilczek90}, $n=4\omega$ and
$\nu=\theta/2\pi$ where $\omega$ is the winding number and $\nu$ the
relative angular momentum in units of $\hbar$ quantized in units of
$\nu+\text{integer}$ in each sector $\alpha$.

In $d\ge 3$ there are only $2$ possible representations of the permutation
group: the one corresponding to the bosonic statistics ($\nu=0$ mod 2) and 
the one corresponding to the fermionic statistics ($\nu=1$ mod 2). 
In $d=2$ one has to choose representations of the braid group
(see chapter 2 of Ref. \cite{Lerda}) and the statistical parameter $\nu$ can be
arbitrary at 
least in principle \footnote{There are restrictions on $\nu$ coming
from the topology of the two dimensional space. For example for
particles moving on a torus (or a 2D box with periodic boundary
conditions), $\nu$ can only be a rational number (see Section
\ref{sec:2d-PB}).}. Particles with 
this property are called {\sl anyons}. In $d=2$ it is not enough to
specify the initial and final configurations to completely characterize the 
system; it is also necessary to specify how the different trajectories wind
or {\sl braid} around each other. In other words the time evolution of 
the particles is important and cannot be neglected in $d=2$. This fact
implies that in order to classify and characterize anyons, the
representations of the permutation group must be replaced by those of
the more complicated {\sl braid group}.   

The following is always true (here $t$ and $t^\prime$ are two
different imaginary times),
\bq
\sum_{i<j}[\theta_{ij}(t^\prime)-\theta_{ij}(t)]=n\pi~~~,
\eq
where the symbol $\theta_{ij}$ denotes the azimuthal angle of particle $j$
with respect to particle $i$ and $n$ is an integer. This can be
interpreted by saying that to complete a loop in configuration space
an integer number of exchanges is always necessary. And one can write
(see chapter 2 of Ref. \cite{Lerda})
\bq \label{thetaij}
\theta_{ij}=\tan^{-1}\left(\frac{x_j^2-x_i^2}{x_j^1-x_i^1}\right)~~~,
\eq
$(x_i^1-x_i^2)$ being the Cartesian coordinates of the $i-th$ particle.

So we can be formally express 
\bq \label{winrep}
\chi(\alpha)=\exp\left[-i\nu\sum_{i,j}\int_0^{\hbar\beta}d\tau\,\frac
{d}{d\tau} \theta_{ij}^{(\alpha)}(\tau)\right],
\eq
Notice that the functions $\theta_{ij}^{(\alpha)}(\tau)$, where
$\alpha$ represents an arbitrary braiding (see chapter 2 of
Ref. \cite{Lerda}) are in
general very complicated and can be specified only when the dynamics
of the particles is fully taken into account. However the formal
definition (\ref{winrep}) may come useful when inserted into the
density matrix expression (\ref{ddm}). 
So that the expression for the diagonal of the density matrix gets 
the suggestive form,
\bq \label{ddm}
\rho(q,q;\beta)&=&\sum_{\alpha\in\pi_1(M_N^2)}\chi(\alpha)
\rho_\alpha(q,q;\beta)\\ \nonumber
&=&\sum_{\alpha\in\pi_1(M_N^2)}
\pint_{q_\alpha(0)=q}^{q_\alpha(\hbar\beta)=q} e^{
-\frac{1}{\hbar}\int_0^{\hbar\beta}d\tau\,\left[{\cal H}(q_\alpha(\tau),
\dot{q_\alpha}(\tau))+i\hbar\nu\sum_{i,j}\frac{d\theta_{ij}^{(\alpha)}
(\tau)}{d\tau}\right]} {\cal D}q_\alpha~.
\eq 

Expression (\ref{ddm}) tells us that instead of dealing with anyons
governed with the Hamiltonian $\cal H$, we can work with bosons whose
dynamics is dictated by the new Hamiltonian ${\cal H^\prime}={\cal
  H}+i\hbar\nu \sum_{i,j}d\theta_{ij}^{(\alpha)}(\tau)/d\tau$. In
particular we could treat fermions governed by an Hamiltonian $\cal H$
as bosons with a ``fictitious'' Hamiltonian ${\cal H^\prime}={\cal
  H}+i\hbar\sum_{i,j}d\theta_{ij}^{(\alpha)}(\tau)/d\tau$. Notice that
this statistical interaction is very peculiar and intrinsically
topological in nature (it is actually a total derivative).  
Its addition to the Hamiltonian ${\cal H}$ does not change the
equations of motion, which are a reflection of the local structure of
the configuration space, but does change the statistical properties of
the particles, which are instead related to the global topological
structure of the configuration space (it can be locally realized as a
gauge theory with a Chern-Simons kinetic term). 

Now since $\rho(q,q;\beta)$ has to be 
a real positive function as well as all the $\rho_\alpha(q,q;\beta)$
one has to add the constraints
\bq
\sum_{\alpha\in\pi_1(M_N^2)}\sin(\nu n_\alpha\pi) \rho_\alpha(q,q;\beta)
=0~~~,\\
\sum_{\alpha\in\pi_1(M_N^2)}\cos(\nu n_\alpha\pi) \rho_\alpha(q,q;\beta)
>0~~~.
\eq

\section{Periodic boundary conditions}
\label{sec:pbc}

The configuration space $M$ of identical hard core two dimensional 
particles has a non trivial topology.
\begin{itemize}
\item If the particles are free to move in $\real^2$ or in 
a finite $L\times L$ box then the configuration space is 
infinitely connected (see fig. \ref{fig:2d}). Its fundamental group is  
the braid group whose representations are labeled by an 
arbitrary parameter $\nu$. This unusual statistics can be implemented
on ordinary particles (for instance bosons) by the addition of a
topological statistical interaction as we saw in Eq. (\ref{ddm}). 
\item If the particles are free to move in a finite box 
with periodic boundary conditions, a torus, a compact Riemannian 
surface of genus 1, then only bosons and fermions are possible
\cite{Lerda} if the multi-particle wavefunctions carry a one
dimensional (appropriate for scalar wave functions) unitary
representation of the braid group. However anyons are possible even on
a torus provided that wave functions with many components are
considered, as for example for spin one-half electrons. In
this case one has to look at higher dimensional representations of the
braid group which lead to the concepts of {\sl generalized fractional
statistics and generalized anyons}
\cite{Einarsson1990,Einarsson1991,Imbo1990,Wen1990}. Now only
fractional statistics are possible and
$\nu=p/q$ can only be a rational number, with $p$ and $q$ coprime
integers and $N=qn$ where $n$ is a non negative integer. This is
essentially due to the requirement to have nonzero winding numbers
along the two periods (the two handles) of the torus: one periodicity
winding acts on a wave function with $k$ components by multiplying all
components by the same phase factor, while the other periodicity
winding mix among themselves the components of the wave function (at
the end of chapter 2 of Ref. \cite{Lerda} the general case of a
Riemannian surface of a generic genus is also made). 
\end{itemize}
In order to avoid periodic boundary conditions one could work on the
surface of a sphere, in this case scalar anyons with fractional
statistics will emerge \cite{Lerda}. 

So this poses the following conceptual problem. If one is to simulate,
for example through the Monte Carlo technique, a system of identical
hard core particles living in two dimensions, he should use, for the
many body wave function of the system contained in a two-dimensional
box of sides $L_1$ and $L_2$, either the Born-von Karman periodic
boundary conditions 
\bq \label{bc}
\psi(\rr_1,\ldots,\rr_j+\LL,\ldots)=
e^{i\Theta/2}\psi(\rr_1,\ldots,\rr_j,\ldots),~~~\forall~j=1,\ldots,N
\eq
with $\Theta=0$ and $\LL=(L_1,L_2)$ or the twisted boundary conditions
\cite{Lin2001}, with $\Theta\neq 0$, to mimic the thermodynamic
limit. Then the fractional statistics or the anyonic nature of the
particles is necessarily changed by the topological change of the
configurational space. Moreover as we will discuss in the conclusions
the twisted boundary conditions, even if they do not alter the
qualitative picture respect to the Born-von Karman boundary
conditions, regarding the topological properties of the underlying
configurational space, they become essential in the description of
anyons or the fractional QHE (see \cite{Lerda} chapter 4). We can in
fact say that in the interchange of two particles each one of the two
changes identity when winding across the boundary (\ref{bc}) as
follows,  
\bq \label{abc}
\psi(\rr_1,\ldots,\rr_j,\ldots,\rr_k,\ldots)=
e^{i\Theta}\psi(\rr_1,\ldots,\rr_k,\ldots,\rr_j,\ldots).
\eq
Since the discovery of the twisted boundary conditions by Chang Lin et
al. in 2001 to optimize the approach to the thermodynamic limit of a
generic Monte Carlo simulation of a many-body system we are unaware of
their use in computer experiment for anyons as in Eq. (\ref{abc}). 

Let us now reduce ourselves to the $N=2$ case. We have seen that when
the particles are free to move on all $\real^d$ then the center of
mass coordinate splits off in a trivial way. Let' s see what we can
easily say about the configuration spaces of particles confined in a
box (B) or in a periodic box (PB). 
We start with a one dimensional space and then study the two
dimensional one.

\subsubsection{For a box in $d=1$ [1d-B]} 
\begin{figure}[htbp]
\begin{center}
\includegraphics[width=8cm]{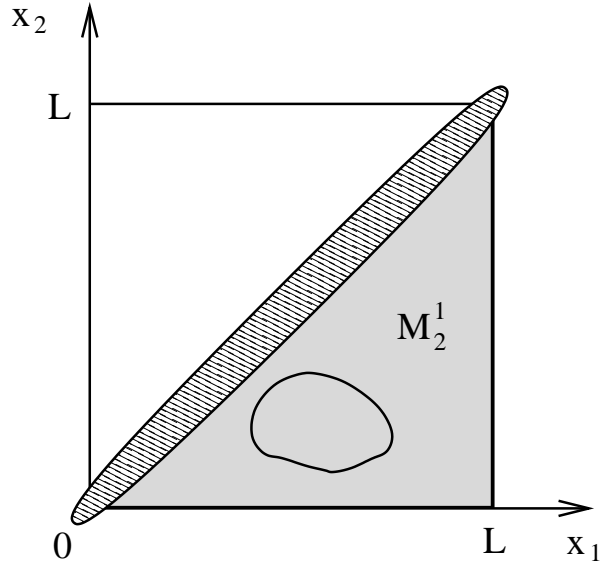}
\end{center}  
\caption{In the plane $(x_1,x_2)$, the uniformly shaded region $M_2^1$
  of Eq. (\ref{M21}) is simply connected. The slashed shaded region is
  the forbidden one.}  
\label{fig:M21}
\end{figure}

Call $x_1\in[0,L]$ and $x_2\in[0,L]$ the particles coordinates. In
this case (see fig. \ref{fig:M21}),  
\bq \label{M21}
M_2^1=\{(x_1,x_2):~~x_2\in[0,L],~x_2<x_1\le L\}~~~,
\eq 
which is simply connected. So only boson statistics is 
allowed.

\begin{figure}[htbp]
\begin{center}
\includegraphics[width=8cm]{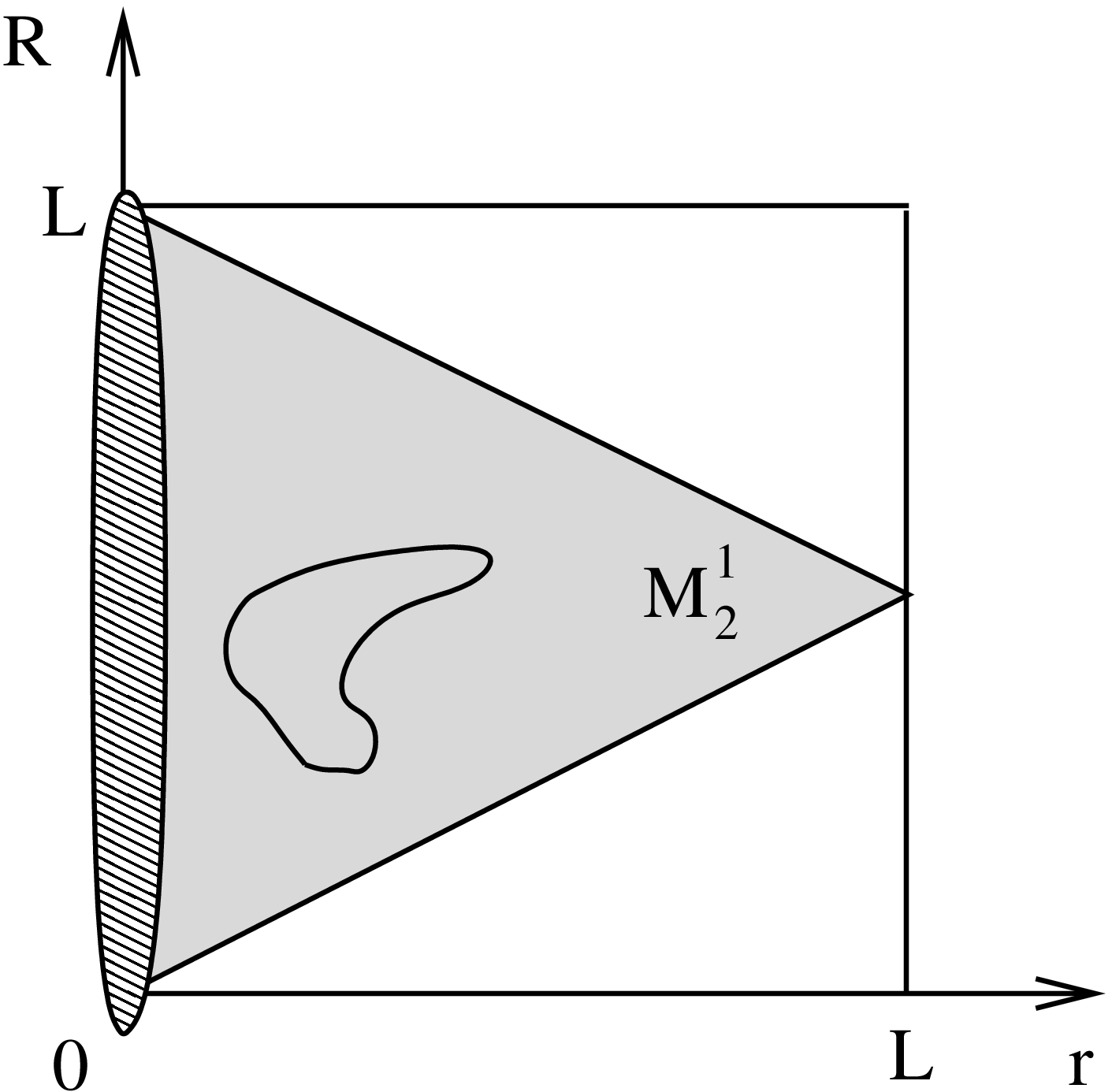}
\end{center}  
\caption{Another view of $M_2^1$ of Eq. (\ref{M21}), now in the plane
  $(r,R)$ with $R$ the center of mass coordinate and $r$ the relative
  coordinate $r$. Again the slashed region is the forbidden one.} 
\label{fig:M21b}
\end{figure}

We could, as well, have introduced the center of mass coordinate
$R=(x_1+x_2)/2\in]0,L[$ and the relative coordinate
$r=x_1-x_2$. Using this coordinates $M_2^1=r_2^1\times]0,L[$ (see
fig. \ref{fig:M21b}). 

As expected again $M_2^1$ is simply connected.

\subsubsection{For a box with periodic boundary conditions in $d=1$ [1d-PB]}
\begin{figure}[htbp]
\begin{center}
\includegraphics[width=10cm]{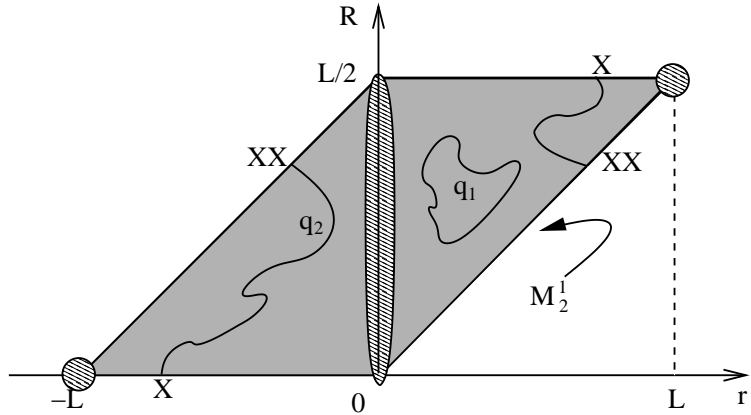}
\end{center}  
\caption{In the plane $(r,R)$ with $R$ the center of mass coordinate
  and $r$ the relative coordinate $r$, we show the uniformly shaded
  region $M_2^1$ of Eq. (\ref{M21b}) which is infinitely
  connected. The points labeled $X$ are the same point. The points
  labeled $XX$ are the same point. The slashed shaded regions are the
  forbidden ones. The path $q_1$ cannot be deformed continuously into
  the path $q_2$.}
\label{fig:M21PB}
\end{figure}

We now consider the case of particles on a circle of length $L$. Using
the center of mass coordinate $R=(x_1+x_2)/2$ and the relative coordinate
$r=x_1-x_2$ one sees by inspection that,
\bq \nonumber
M_2^1=\{(r,R):~~R\in[0,L/2],~2R-L\le r\le 2R,\\ \nonumber
(2R,R)=(2R-L,R),~
(-r,0)=(r,L/2)\}\\ \label{M21b}
-\{(0,R)~\forall R,~(-L,0),~(0,L/2)\}~~~,
\eq
which is infinitely connected (as shown in fig. \ref{fig:M21PB} two
loops with different winding around the missing point $(-L,0)=(L,L/2)$
are homotopically inequivalent). So anyons with arbitrary statistics
$\nu$ is allowed.

\begin{figure}[htbp]
\begin{center}
\includegraphics[width=12cm]{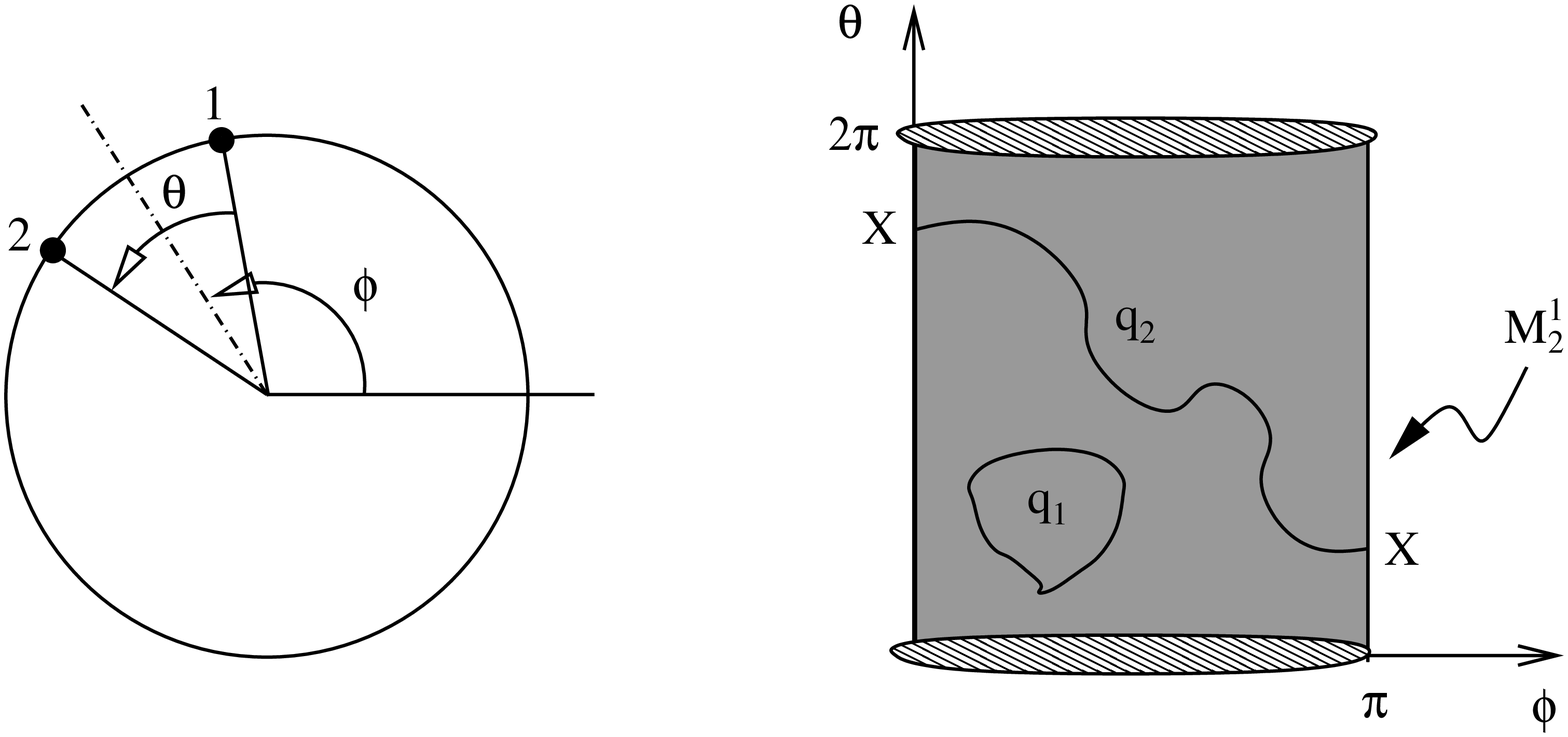}
\end{center}  
\caption{On the right we show the meaning of the angles $\theta$,
  between the two particles 1 and 2 on the circle, and $\phi$, the
  polar angle to the line joining the center of the circle to the
  center of mass of the two particles. The uniformly shaded region
  $M_2^1$ of Eq. (\ref{M21b}) is infinitely connected. The points labeled
  $X$ are coincident. The slashed shaded regions are the forbidden
  ones. The path $q_1$ cannot be deformed continuously into the path
  $q_2$.}  
\label{fig:M21PBb}
\end{figure}

The same thing can be seen introducing the center of mass angle 
$\phi$ and the relative angle $\theta$ (see fig. \ref{fig:M21PBb}).
The rectangle in the $(\phi,\theta)$ plane defined by $0\le\phi\le\pi$
and $0\le\theta\le2\pi$ includes all possible configurations, except
for the left and right edges where $(0,\theta)$ and $(\pi,2\pi-\theta)$
both represent the same configuration. Because of this identification the 
rectangle becomes a M\"obious band which is still infinitely
connected. In this case though even with a multi-component
wave function the statistics must remain arbitrary and not fractionary
as in the two dimensional one since we have only one periodicity.

\subsubsection{For a box in $d=2$ [2d-B]} 
Using the same argument used for the [1d-B] we can say
that $M_2^2=r_2^2\times[0,L]^2$ where $r_2^2$ is a space 
with the same topology as the cone without the tip introduced in the
case of particles without boundaries. The only difference being that
the cone now does not extend to infinity but is finite and its height
depends on $L$. So once again, since $r_2^2$ is infinitely connected,
also $M_2^2$ is. And anyon statistics is allowed with arbitrary $\nu$.

\subsubsection{For a box with periodic boundary conditions in $d=2$ [2d-PB]}
\label{sec:2d-PB}
In this case we can say that something similar was happening from
going from the 2d plane ($M$ infinitely connected) to the 3d space
($M$ doubly connected). Now in [1d-PB] M is infinitely connected and
in [2d-PB] $M$ is doubly connected. We split again $M_2^2$ into the
product of the center of mass configuration space and of the
two impenetrable particles relative coordinates one, $r_2^2$. It turns
out that now, due to the periodic boundary conditions, $r_2^2$ is a cone
without the tip, of finite height, as in Fig. \ref{fig:2d}, and with
the end points of a diameter of the base identified. This is a doubly
connected space. All this is only true if we consider scalar wave
functions, i.e. one dimensional representations of the fundamental
group of the configuration space. For wave functions with many
components the generators of the representations of the fundamental
group of the configuration space are such that \cite{Lerda} $\nu=p/q$ a 
rational number, with $p$ and $q$ coprime numbers and a restriction on
the total number of particles, $N=qn$, where $n$ is a non negative
integer. For an extensive discussion of anyons on compact surfaces and
on the torus in particular, we refer the reader to the review by
R. Iengo and K. Lechner \cite{Iengo1992}. 

\section{Conclusions} 
\label{sec:conclusions} 
Twisted boundary conditions play a relevant role in the anyons problem
where the topology of the underlying configuration space determines
the statistics of the particles. We review various cases. For scalar
many body  
wave functions on the segment or the infinite line one can have only
bosons, on the circle one can only have anyons with arbitrary
statistics, on the square or the infinite plane one can also have only
anyons with arbitrary statistics, and on the torus which has two
periodicities only bosons and fermions are allowed as on the infinite
three dimensional Euclidean space. We gave a proof of these different
behaviors for just a two-body system. This is enough to determine the
anyonic symmetry of the many-body wave function as we discussed in
Section \ref{sec:smp} but one cannot exclude other kinds of three and
higher body symmetries where it is necessary to substitute
$\theta_{ij}$ of Eq. (\ref{thetaij}) with a different
$\theta_{ijk\ldots}$. We gave proofs of these circumstances based on
the geometrical topological properties of the configurational space in
each case, which we regard as the simplest way to proceed.

If we allow for a many components wave function on the
torus we may have anyons {\sl but} with only {\sl fractional
  statistics} which proved to give an interpretation for the fractional
QHE. In this case a series of new states of matter emerge as
incompressible quantum liquids \cite{Laughlin1983a,Laughlin1983b}
around which the low-energy excitations are localized quasi-particles
with unusual fractional quantum numbers, i.e. anyons. The Laughlin
variational ground-state wave functions requires the statistics,
$\nu$, to be an odd
integer $m$ whereas the excited states require it to be
rational. Laughlin chooses the trial ground-state wave function of the
Bijl-Dingle-Jastrow product form 
\bq
\psi_m={\cal N}_m\prod_{i<j}(z_i-z_j)^m e^{-\frac{1}{4\ell_0^2}\sum_i|z_i|^2},
\eq
where ${\ell_0=\sqrt{\hbar c/eB}}$ is the magnetic length, $B$ the
magnetic field orthogonal to the metallic plate, $z_i$ is
the complex coordinate of the i-th electron and ${\cal 
  N}_m$ is a normalization factor. Since $m$ is an odd integer, $\psi$
is totally antisymmetric, and so it describes ordinary fermions. 
The prefactor $(z_i-z_j)^m$ is also of the Jastrow type: it has a zero
of order $m$ at coincident points $(z_i = z_j)$, showing that
electrons tend very strongly to repel each other in a way that is
appropriate to minimize the Coulomb interaction. If $z_i$ goes around
$z_j$ by an angle $\Delta\theta$ the wave function acquires a phase
$e^{im\Delta\theta}$, as if each particle carried $m$ units of
flux. This allows Laughlin to use the fact that the $|\psi_m|^2$ can
be interpreted as the Boltzmann factor $e^{-\beta \varphi}$ of a One
Component Plasma of classical particles of charge $Q=m$ living
in two dimensions where the neutralizing background has a surface
charge density $\sigma=m/2\pi\ell_0^2$ at an inverse temperature
$\beta=2/m$. The coupling constant of the plasma is $\Gamma=\beta
Q^2=2m^2$ and its properties are available exactly analytically at the
special value of the coupling constant $\Gamma=2$
\cite{Jancovici81b,Fantoni03jsp,Fantoni08c} when the two dimensional  
electron gas corresponds to a full Landau level $m=1$ (see
Ref. \cite{Lerda} chapter 8).    

A word of caution when thinking at the physical implications of all
this are nonetheless necessary. From a purely conceptual point of view
the fact that in order to have a fractional statistics one has to
impose twisted periodic boundary conditions that are an 
artificial means to approach the thermodynamic limit and have no
physical meaning sheds some doubts on the relevance of the anyonic
theory on the interpretation of the fractional QHE. From the point of
view of the numerical experiment the presence of a magnetic field
implies that the ground state wave function will, in general, be
complex valued and in order to deal with the symmetry given by the
anyonic statistics one should use methods similar to the ones used in
Ref. \cite{Zhang1993,Jones1997}. Also we proposed to combine these
methods with the twisted boundary conditions first employed in 2001 by
Chang Lin et al. \cite{Lin2001} for a generic many-body system.
It would be desirable to perform the simulation on a sphere with a
Dirac magnetic monopole at the center \cite{Melik2001} in order to be
able to simulate scalar anyons with fractional statistics, without the
necessity of implementing any sort of boundary conditions.

Another issue in disfavor of the description of the physically
observed QHE is the fact that in a laboratory the electrons will
surely not be exactly living in a two dimensional world but one deals
rather with a quasi two dimensional, very very thin, metallic layer
\cite{Fantoni-tesi-laurea} at the interface between two different
semiconductors or between a semiconductor and an insulator even if the
low temperature and the strong magnetic field freeze the motion along
the direction perpendicular to the layer (something similar as
explained in the satirical novella by the English schoolmaster Edwin
Abbott Abbott: ``Flatland: A Romance of Many Dimensions'' first
published in 1884 by Seeley \& Co. of London). This of course would
modify also the Coulomb potential of interaction between the electrons
from one $\propto-\log(r/L)$ to one $\propto 1/r$, with $r$ the
separation between electrons, which are in any case both divergent at
$r=0$. Naturally the Coulomb repulsion is essential to give the
incompressibility condition avoiding two particles to overlap. 

The real experiment is too complicated to describe in its completeness
so one has to resort to approximations and the approximation of
considering the electrons as ``living'' in a two dimensional world
with periodic twisted boundary conditions seems to be an effective
one. There are many experiments in the field. One I am most interested
in is Ref. \cite{Salamon1997} where it is shown that the sign
of the Hall effect in the transport properties of doped lanthanum
manganites films for small polaron \cite{Fantoni12d,Fantoni13a}
hopping can be ``anomalous''. A small 
polaron based on an electron can be deflected in a magnetic field as
if it were positively charged and, conversely, a hole-based polaron
can be deflected in the sense of a free electron. Measurements of the
high-temperature Hall coefficient of manganite samples reveal that it
exhibits Arrhenius behavior and a sign anomaly relative to both the
nominal doping and the thermoelectric power. The results are discussed
in terms of an extension of the Emin-Holstein theory of the Hall
mobility in the adiabatic limit.

There are now several proposed experiments
aimed at identifying the existence of non-Abelian
statistics in nature. Non-Abelian phases are gapped phases
of matter in which the adiabatic transport of one excitation
around another implies a unitary transformation
within a subspace of degenerate wavefunctions which differ
from each other only globally \cite{Read1992}.

Another more recent experimental interest in anyons is for topological
quantum computation \cite{Stern2010,Keyserlingk2015}: Systems
exhibiting non-Abelian statistics can store topogically protected
qubits \cite{DasSarma2005}.


\begin{acknowledgments} 
I would like to acknowledge fruitful discussions with Rob Leigh,
Eduardo Fradkin, Michael Stone, and last but not least Myron Salamon
who showed me the physics of calorimeters, way back in 2000 in Urbana.
\end{acknowledgments} 
\bibliography{anyons-pbc}

\end{document}